\documentclass[prb,twocolumn]{revtex4}%

\usepackage{graphicx}
\usepackage{amsmath}
\usepackage{ulem}
\usepackage{color}

\newcommand{\be}{\begin{equation}}
\newcommand{\ee}{\end{equation}}
\newcommand{\bea}{\begin{eqnarray*}}
\newcommand{\eea}{\end{eqnarray*}}
\newcommand{\bean}{\begin{eqnarray}}
\newcommand{\eean}{\end{eqnarray}}

\begin{document}

\draft
\title
{\bf  Optimizing thermoelectric efficiency of superlattice
nanowires at room temperature}

\author{David M T Kuo$^{1}$, C. C. Chen,$^{2}$  and Yia-Chung Chang$^{2,3}$}

%\author{David M T Kuo}

\address{$^{1}$Department of Electrical Engineering and Department of Physics, National Central
University, Chungli, 320 Taiwan}

\address{$^{2}$Research Center for Applied Sciences, Academic Sinica,
Taipei, 11529 Taiwan}

\affiliation{$^3$ Department of Physics, National Cheng Kung
University, Tainan, 701 Taiwan}

\date{\today}

\begin{abstract}
It is known that the figure of merit ($ZT$) of thin nanowires can
be significantly enhanced at room temperature due to the reduction
of phonon thermal conductance arising from the increase of
boundary scattering of phonons. It is expected that the phonon
thermal conductance of nanowires filled with quantum dots (QDs)
will be further reduced. Here we consider a superlattice nanowire
(SLNW) modeled by a linear chain of strongly coupled QDs connected
to electrodes. We study the dependence of $ZT$ on the QD energy
level ($E_0$) (relative to the Fermi level $E_F$ in the
electrodes), inter-dot coupling strength ($t_c$), tunneling rate
($\Gamma$), and temperature $T$ in order to optimize the design.
It is found that at room temperature the maximum power factor
occurs  when $(E_0-E_F)/k_BT \approx 2.4$ and $\Gamma=t_c$, a
result almost independent of the number of QDs in SLNW as long as
$t_c/k_BT <0.5$. By using reasonable physical parameters we show
that thin SLNW with cross-sectional width near $3~nm$ has a
potential to achieve $ZT\ge3$.
\end{abstract}

\maketitle
\section{Introduction}
Extensive studies have shown that in bulk thermoelectric
materials, it is difficult to achieve a figure of merit ($ZT$)
larger than one at room temperature.[1,2] With the advances of
nanostructure technology, many experiments nowadays can realize
$ZT$ larger than one at room temperature in low-dimensional
structures.[3,4] The search of nanostructured materials with
significantly improved $ZT$ is still a subject of hot pursuit. If
a material with $ZT \ge 3$  at room temperature (which corresponds
to a Carnot efficiency around 30\% [4] can be found, it will
brighten the scenario of thermoelectric devices tremendously.[1,2]
For example, thermoelectric generators (TEGs) using human body as
a heat source can be applied to wearable electrical powers, which
are very useful for commercial wireless communication and low
power electronics[5]. Thermoelectric coolers will also become a
viable option for many applications.[1,2] {It has been predicted
theoretically  that  $ ZT \ge 3 $ can be achieved in thin
semiconductor nanowires.[6,7] However, no experimental realization
of such impressive TE devices has been reported.[1,2]

The finding of $ZT=1$ in silicon nanowires at room temperature[8]
has inspired further studies of thermoelectric properties of
silicon-based nanowires because of the advantages of low cost and
the availability of matured fabrication technology in silicon
industry.[8-10] Whether $ZT \ge 3$ exists in silicon-based
nanowires at room temperature becomes an interesting topic. Monte
Carlo simulations[11] have demonstrated that in heavily-doped Si
naowires $ZT$ does not increase dramatically with decreasing wire
cross section since the electron conductance suffers stronger
ionized-impurity scattering as the size reduces. Thus, to improve
$ZT$ it is better to use intrinsic nanaowires. Silicon nanowire
filled with quantum dots (QDs) may provide an alternative means to
realize the values of $ZT \ge 1$.[1,2] Furthermore, it is
estimated that Si/Ge superlattice nanowire (SLNW) with an
optimized period around 5nm and cross-sectional area around $3nm
\times 3nm$ can lead to an reduction of phonon thermal conductance
by one order of magnitude in comparison with pristine Si
nanowires.[12,13] Therefore, it is desirable to study the
dependence of $ZT$ on relevant physical parameters of Si/Ge SLNWs
near room temperature.

\begin{figure}[h]
\centering
\includegraphics[trim=0cm 0cm 0cm 0cm,clip,angle=-90,scale=0.3]{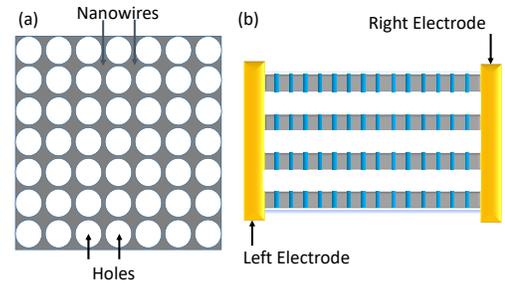}
\caption{(a) Schematic sketch of the creation of superlattice
nanowires (SLNWs). (b) Side view of SLNWs connected to
electrodes.}
\end{figure}

Here, we perform theoretical calculations of the thermoelectric
properties of intrinsic SLNW connected with electrodes by using a
linear chain of strongly coupled QDs. The electron carriers in
SLNWs are provided by  metallic electrodes with Fermi level below
the conduction band minimum as proposed in[14]. The SLNW structure
can be realized by starting with a superlattice with optimized
period to minimize the phonon thermal conductance. The creation of
nanowires may be achieved by lithographically drilling a periodic
array of closely-spaced holes[10], leaving an array of nanowires
(with a star-shaped cross-section) as depicted in Fig.~1. To
reduce the electron scattering from the rough surface, the side
surfaces of the nanowires are oxidized to form a high electron
barrier, which serves the purpose of confining electrons in the
the central region of the star-shaped nanowires. Due to the strong
confinement of electrons, the electron-surface scattering can be
reduced. At the same time the acoustic phonons are not so well
confined in the central region which will then suffer more surface
scattering. With such a design, we found that $ZT \ge 3 $ may be
achieved in a thin SLNW with reasonable physical parameters
adopted. Our theoretical studies should provide a useful guideline
for the design of future thermoelectric devices operating at room
temperature.

\section{Formalism}
To model the thermoelectric properties of SLNWs, we consider a
linear chain of strongly-coupled QDs with a system Hamiltonian
given by an Anderson model $H=H_0+H_{QD}$,[15] where
\begin{eqnarray}
H_0& = &\sum_{k,\sigma} \epsilon_k
a^{\dagger}_{k,\sigma}a_{k,\sigma}+ \sum_{k,\sigma} \epsilon_k
b^{\dagger}_{k,\sigma}b_{k,\sigma}\\ \nonumber &+&\sum_{k,\sigma}
V^L_{k,L}d^{\dagger}_{L,\sigma}a_{k,\sigma}
+\sum_{k,\sigma}V^R_{k,R}d^{\dagger}_{R,\sigma}b_{k,\sigma}+c.c.
\end{eqnarray}
The first two terms of Eq.~(1) describe the free electron gas in
the left and right electrodes. $a^{\dagger}_{k,\sigma}$
($b^{\dagger}_{k,\sigma}$) creates  an electron of momentum $k$
and spin $\sigma$ with energy $\epsilon_k$ in the left (right)
electrode. $V^L_{k,L}$ ($V^R_{k,R}$) describes the coupling
between the left (right) lead with its adjacent QD.
$d^{\dagger}_{L(R),\sigma}$ ($d_{L(R),\sigma}$) creates (destroys)
an electron in the QD connected to the left (right) lead.
\begin{small}
\begin{equation}
H_{QD}= \sum_{\ell,\sigma} E_{\ell}
d^{\dagger}_{\ell,\sigma}d_{\ell,\sigma}+ \sum_{\ell \neq j}
t_{\ell,j} d^{\dagger}_{\ell,\sigma} d_{j,\sigma},
\end{equation}
\end{small}
where { $E_{\ell}$} is the QD energy level in the ${\ell}$-th QD
and $t_{\ell,j}$ describes the electron hopping strength between
the ${\ell}$-th and $j$-th QDs.  Here, for simplicity, we consider
only one energy level for each QD, which is suitable for nanoscale
QDs with no vally degeracy. For example, a cylidrical GaAs QD with
high lateral potential barrier resulting from oxidation and
vertical confinement via the band onset between GaAs and AlGaAs
has a bound state (which is well separated in energy from the
excited states) for diameter naer $3~nm$ and height near
$5~nm$.[16] For silicon QDs in the Si/Ge SLNW, we expect the
valley degeneracy to lead to enhanced density of states in the
energy range of interest, which can actually improve ZT
further.[17] It should be noted that the Hubbard-like terms for
Coulomb interactions between electrons in the SLNWs are neglected,
since the electrons are delocalized along the transport direction
in the strong hopping limit considered.

To study the transport properties of QDs junction connected with
electrodes, it is convenient to use the Green-function technique.
The electron and heat currents from reservoir $\alpha$ to its
adjacent QD are calculated according to the Meir-Wingreen
formula[18]
\begin{eqnarray}
J^n_\alpha &=&\frac{ie}{h}\sum_{j\sigma}\int
{d\epsilon}(\frac{\epsilon-\mu_{\alpha}}{e})^{n}
\Gamma_{\alpha}(\epsilon) [ G^{<}_{\alpha,\sigma} (\epsilon)\\
\nonumber &+& f_\alpha (\epsilon)( G^{r}_{\alpha,\sigma}(\epsilon)
- G^{a}_{\alpha,\sigma}(\epsilon) ) ],
\end{eqnarray}
where $n=0$ is for the electrical current and $n=1$  for the heat
current.
$\Gamma_{L(R)}(\epsilon)=\sum_{k}|V_{k,L(R)}|^2\delta(\epsilon-\epsilon_k)$
is the tunneling rate for electrons from the left (right)
reservoir and entering the left (right) QD.
$f_{\alpha}(\epsilon)=1/\{\exp[(\epsilon-\mu_{\alpha})/k_BT_{\alpha}]+1\}$
denotes the Fermi distribution function for the $\alpha$-th
electrode, where $\mu_\alpha$  and $T_{\alpha}$ are the chemical
potential and the temperature of the $\alpha$ electrode. $e$, $h$,
and $k_B$ denote the electron charge, the Planck's constant, and
the Boltzmann constant, respectively. $G^{<}_{\alpha,\sigma}
(\epsilon)$, $G^{r}_{\alpha,\sigma}(\epsilon)$, and
$G^{a}_{\alpha,\sigma}(\epsilon)$ denote the frequency-domain
representations of the one-particle lessor, retarded, and advanced
Green's functions, respectively.

{In the linear response regime electrical conductance ($G_e$),
Seebeck coefficient ($S$) and electron thermal conductance
($\kappa_e$) can be evaluated by using Eq.~(3) with a small
applied bias $\Delta V=(\mu_L-\mu_R)/e$ and temperature difference
across junction $\Delta T=T_L-T_R$.[17] For simplicity, we
calculate these thermoelectric coefficients in terms of Landauer
formula[19,20]. Here we have $G_e=e^2{\cal L}_{0}$, $S=-{\cal
L}_{1}/(eT{\cal L}_{0})$ and $\kappa_e=\frac{1}{T}({\cal
L}_{2}-{\cal L}^2_{1}/{\cal L}_{0})$. Thermoelectric coefficients
${\cal L}_n$ are given by
\begin{equation}
{\cal L}_n=\frac{2}{h}\int d\epsilon {\cal
T}_{LR}(\epsilon)(\epsilon-E_F)^n\frac{\partial
f(\epsilon)}{\partial E_F},
\end{equation}
where $f(\epsilon)=1/(exp^{(\epsilon-E_F)/k_BT}+1)$ is the Fermi
distribution function of electrodes at equilibrium temperature $T$
and ${\cal T}_{LR}(\epsilon)$ is the transmission coefficient.
$E_F$ is the Fermi energy of electrodes. The expression of ${\cal
T}_{LR}(\epsilon)$ in Eq. (4) is given by [20,21]
\begin{small}
\begin{equation}
{\cal T}_{LR}(\epsilon)=\frac{-4 \Gamma_L
\Gamma^{eff}_R(\epsilon)}{\Gamma_L+\Gamma^{eff}_R(\epsilon)}
Im(G^r_{L}(\epsilon)),
\end{equation}
\end{small}
where
$G^r_{L}(\epsilon)=1/(\epsilon-E_1+i\Gamma_L-\Sigma_{1,N}(\epsilon))$
denotes the one-particle retarded Green function of the leftmost
QD with the energy level of $E_1$, in which the self energy
$\Sigma_{1,N}(\epsilon)$ resulting from electron tunneling from
the leftmost QD to the right electrode mediated by N-1 QDs is
given by [21]
\begin{small}
\begin{eqnarray}
\Sigma_{1,N}(\epsilon)
&=&\frac{t^2_{1,2}}{\epsilon-E_2-\frac{t^2_{2,3}}{\epsilon-E_3-......\frac{t^2_{N-1,N}}{\epsilon-E_N+i\Gamma_R}}},
\end{eqnarray}
\end{small}
where $N$ denotes the total number of QDs. The rightmost QD is the
$N$th QD. The effective tunneling rate $\Gamma^{eff}_R =
-Im(\Sigma_{1,N}(\epsilon))$. For simplicity, we assume
$E_{\ell}=E_0$ and $t_{\ell,j}=t_c$ for all $\ell$ and $j$ being
the nearest neighbor of $\ell$, and $\Gamma_L=\Gamma_R\equiv
\Gamma$.

The figure of merit $ZT=S^2G_eT/(\kappa_e+\kappa_{ph})$ contains
the phonon thermal conductance ($\kappa_{ph}$) of SLNW, which can
not be neglected at room temperature.[17] Many theoretical efforts
have been devoted to the study of phonon thermal conductance of
silicon nanowires.[22,23] Here, we adopt the formula of phonon
thermal conductance as given in Ref. [23], which can well describe
the experimental results of Si nanowires.[23,24] We have
\begin{equation} \kappa_{ph,0}(T)=\frac 1 h \int d\omega {\cal T}(\omega)_{ph}
\frac {\hbar^3 \omega^2}{k_B T^2}\frac
{e^{\hbar\omega/k_BT}}{(e^{\hbar\omega/k_BT}-1)^2}, \label{phC}
\end{equation}
where $\omega$ is the phonon frequency and ${\cal T}_{ph}(\omega)$
the throughput function. In [12,13,25] it is theoretically
demonstrated that $\kappa_{ph}(T)$ of Si nanowires filled with QDs
is significantly reduced from the value $\kappa_{ph,0}(T)$ for Si
nanowires without QDs for a wide range of temperatures. To
simulate the phonon thermal conductance of SLNWs (which is filled
with QDs), we use a simple expression \( \kappa_{ph}(T)=F_s
\kappa_{ph,0}(T), \) where $F_s$ is a dimensionless scaling factor
used to describe the phonon scattering resulting from the
interface scattering of QDs, a filtering mechanism for short
wavelength acoustic phonons.[12,13] Values of $F_s$ can range from
0.1 to 1.[12,13,25] It is well known that QD size fluctuation and
surface roughness will also degrade the electron transport[26,27]
For simplicity, we assume that the electrical conductance and
thermal conductance due to electron-phonon scattering and electron
scattering from the surface roughness and QD size fluctuation can
be modeled by a same scaling factor $F_e=1/(1+L/\lambda)$, where
$L$ and $\lambda$ denote the channel length and electron mean free
path, respectively. Namely, the electrical conductance and thermal
conductance for realistic SLNWs are expressed as $G'_e=F_e G_e$
and $\kappa'_e=F_e \kappa_e$, respectively. The resulting $ZT$ now
reads \( ZT=S^2G_eT/(\kappa_e+\kappa_{ph,0}f_s), \) where
$f_s=F_s/F_e$. For  an optimized design of SLNW as  illustrated in
Fig. 1, it is conceivable that $f_s$ can be less than 1. $F_e$ is
roughly proportional to the ratio of mean-free path to the channel
length. The mean-free path of electrons in germanium/silicon
nanowires reported is beyond 170nm at room temperature.[27] Thus,
$F_e \approx 0.2$ for a micron size SLNW considered here, and the
best value of $f_s$ can be expected is around  0.5. Taking the
possible size nonuniformity of QDs into account, we shall consider
the value of $f_s$ to be between 0.7 and 1 in our simulation and
examine what values of $ZT$ can be achieved for SLNWs.

\section{Results and discussion}
Before discussing the optimization of $ZT$ at room temperature, we
first study the power factor ($PF=S^2G_e$) of SLNWs as a function
of system parameters such as QD energy level (relative to $E_F$)
and the tunneling rate ($\Gamma$), since the behavior of $ZT$  is
mostly determined by $PF$ in the limit $\kappa_{ph}/\kappa_e \gg
1$. Figure 2 shows the electrical conductance, Seebeck coefficient
and power factor as functions of QD energy level ($E_0$) relative
to the Fermi level ($E_F$)   at two different temperatures.  All
energy scales are in units of $\Gamma_0$, which is taken to be
$1~meV$. In the strong-hopping limit, $t_c/\Gamma \gg 1$, the
expression of ${\cal T}_{LR}(\epsilon)$ can be approximated by
\begin{small}
\begin{equation}
{\cal T}_{LR}(\epsilon)=\Pi^{N}_{n} \frac{4\Gamma^2
(t^2_c)^{N-1}}{(\epsilon-[E_0-2t_c
\cos(\frac{n\pi}{N+1})])^2+(\Gamma_n)^2},
\end{equation}
\end{small}
where $\Gamma_n=\Gamma \sqrt{2/(N+1)}\sin(\frac{n\pi}{N+1})$ for
$n=1,\cdots,N$. At low temperature ($k_BT=1\Gamma_0$), the
behavior of $G_e$ can be roughly described by ${\cal
T}_{LR}(E_F)$, which is nonzero only when $E_F$ falls within the
SLNW band, which covers the range from -32 to 32$\Gamma_0$, since
the band width is $4t_c=64\Gamma_0$. [See Fig. 2(a)] The  $G_e$
spectrum becomes significantly broadened  near room temperature
with $k_BT=25\Gamma_0$. The enhancement of $G_e$ outside the SLNW
band is due to the thermionic effect.[14] As seen in Fig. 2(b),
the Seebeck coefficient $S$ is an antisymmetric function of
$\Delta=E_0-E_F$. Such a bipolar behavior has been reported in
several studies.[19,21] The negative sign of $S$ indicates the
main contribution coming from electrons tunnel through the
resonant channels of the band above the Fermi level, while  the
positive sign of $S$ indicates the main contribution coming from
holes tunneling through the band below $E_F$. As seen in Fig. 2(c)
the $PF$ spectrum shows two strong peaks when $E_F$ is near the
SLNW band edges at low temperature, and the peaks are broadened
and shifted away from the band region as temperature increases.

\begin{figure}[h]
\centering
\includegraphics[angle=-90,scale=0.3]{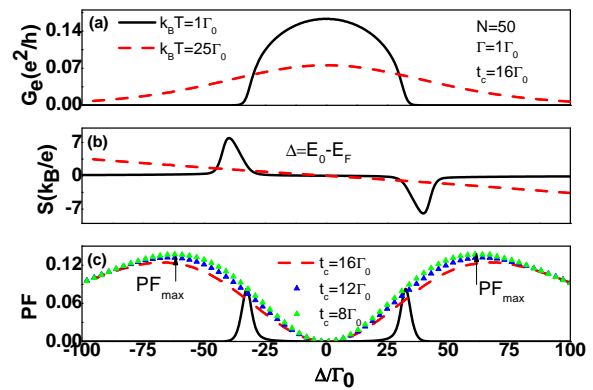}
\caption{(a) Electrical conductance, (b) Seebeck coefficient and
(c) power factor ($PF$) as a function of $\Delta=E_0-E_F$ for two
different temperatures in an SLNW junction with electron hopping
strengths $t_{c}=16\Gamma_0$.  In (c) we also add results for
$t_c=8$  and $ 12\Gamma_0$ (triangles) at $k_BT=25\Gamma_0$. Other
parameters used are $\Gamma_L=\Gamma_R=\Gamma=1\Gamma_0$ and
$N=50$.}
\end{figure}

To find the maximum of $PF$ at room temperature
($k_BT=25\Gamma_0$), we also calculate  $PF$ as a function of
$\Delta$ for various coupling strengths. It is found that the
maximum $PF$ occurs near $\Delta=60\Gamma_0=2.4k_BT$.
Interestingly, this feature is nearly independent of the number of
QDs in SLNW as long as $t_c/k_BT$ is small.

Let's consider the case of $N=2$, which has an analytical
solution. We have
\begin{small}
\begin{equation}
{\cal T}_{LR}(\epsilon)=\frac{4\Gamma_L\Gamma_R
t^2_{c}}{|(\epsilon-E_0+i\Gamma_L)(\epsilon-E_0+i\Gamma_R)-t^2_{c}|^2}.
\label{TLR}
\end{equation}
\end{small}
Under the assumption of $\Gamma/k_BT \ll t_c/k_BT \ll 1$, we have
\begin{small}
\begin{equation}
{\cal L}_0=\frac{1}{hk_BT}\frac{\pi\Gamma~t^2_{c}
}{2(t^2_{c}+(\Gamma/2)^2)} \frac{1}{cosh^2(\frac{\Delta}{2k_BT})}
\end{equation}
\end{small}
and
\begin{small}
\begin{equation}
{\cal L}_1=\frac{1}{hk_BT}\frac{\pi\Gamma~t^2_{c}
}{2(t^2_{c}+(\Gamma/2)^2)}
\frac{\Delta}{cosh^2(\frac{\Delta}{2k_BT})}.
\end{equation}
\end{small}
From Eqs.~(10) and (11), we get $S=-\frac{\Delta}{eT}$ which
explains  the linear behavior of $S$ at $k_BT=25\Gamma_0$ in Fig.
2(b). Furthermore,
\begin{small}
\begin{equation}
PF= \frac{(\Delta/eT)^2
G_0}{k_BT}\frac{\pi\Gamma~t^2_{c}}{2(t^2_{c}+(\Gamma/2)^2)}
\frac{1}{cosh^2(\frac{\Delta}{2k_BT})},
\end{equation}
\end{small}
where $G_0=e^2/h$ denotes the electron quantum conductance. The
above equation gives a maximum $PF$ at $\Delta/k_BT=2.4$. This
result holds well for any number of $N$ (tested up to $N=100$) as
long as $t_c/k_BT<0.5$.

In Fig. 2, we have considered the case with $\Gamma=1\Gamma_0$.
Next, we study how the thermoelectric properties of SLNW are
affected by increasing the tunneling rate $\Gamma$. Figure 3 shows
the calculated $G_e$, $S$ and $PF$ as functions of $\Gamma$ for
various values of $t_c$ at the optimum condition with
$\Delta=2.4k_BT=60\Gamma_0$. The $\Gamma$ dependence of $PF$  is
dominated by $G_e$ since $S$ is  almost independent of $\Gamma$ as
can be seen in Fig.~3(b). It is found that the maximum $PF$ occurs
when $\Gamma=t_c$, as indicated by arrows in Fig.~3(c). We found
this relation also holds approximately for any value of $N$.  For
comparison, we also add the results obtained by Eq.~(\ref{TLR})
for $N=2$ (curves with triangle.[28]

%Again, we found this relation holds approximately for any value of
%$N$, starting with $N=2$. For the $N=2$ case, we obtain \be G_e=
%\frac{e^2}{h}\frac{4\Gamma^2t^2_{c} }{t^2_{c}+\Gamma^2} f(T). \ee
%where $f(T)$ is independent of $\Gamma$. The above expression
%gives a maximum $G_e$ at $\Gamma=t_c$.

\begin{figure}[h]
\centering
\includegraphics[angle=-90,scale=0.3]{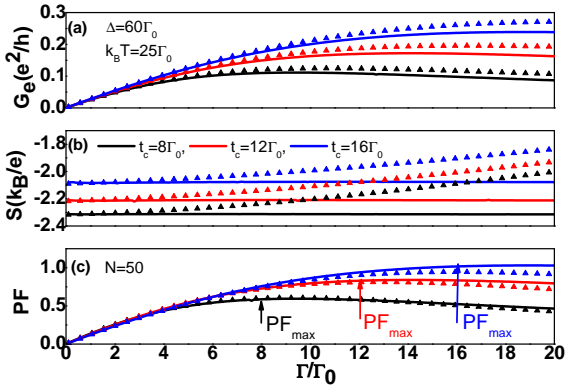}
\caption{(a) Electrical conductance  $G_e$, (b) Seebeck
coefficient $S$, and (c) power factor ($PF$) as a function of
electron tunneling rate $\Gamma$ for various $t_c$  at
$k_BT=25\Gamma_0$.  Solid curves are for $N=50$. Triangles are
obtained by Eq.~(\ref{TLR}) for N=2.Other physical parameters are
the same as those of Fig.~2.}
\end{figure}

Although previous theoretical studies have predicted high $ZT$
values of QD junctions in the Coulomb blockade regime [19]for the
case of $\kappa_e\gg \kappa_{ph}$, it is valid only at extremely
low temperature. To illustrate the importance of $\kappa_{ph}$, we
show in Fig. 4 $\kappa_e$ and $\kappa_{ph,0}$ as functions of
temperature at $\Delta=60~\Gamma_0$ and $\Gamma=t_c$. As seen in
Fig. 4(a) $\kappa_e$  reaches a maximum at $300~K$ and the maximum
value is enhanced with increasing $t_c $ or $\Gamma$. Here we have
adopted the optimum case with $\Gamma=t_c$. For the Si/Ge
superlattice with an optimized period of $5~nm$, the SLNW with
$N=200$ has a length of $L=1000~nm$. Therefore, we are interested
in $\kappa_{ph}$ of nanowires with length of $L=1000~nm$. The
calculated $\kappa_{ph,0}$ based on Eq.~7 for silicon nanowires
with surface roughness width $\delta=2~nm$ and $L=1000~nm$ are
shown in Fig. 4(b) for three different diameters ($D=3, 4, 5nm$).
The magnitude of $\kappa_{ph}$ is reduced very quickly when $D$ is
reduced. We find that $\kappa_{ph,0}$ is larger than $\kappa_{e}$
at $T=300K$ even though for the $D=3~nm$. Both  $\kappa_e$ and
$\kappa_{ph}$ will be included in the calculation of $ZT$.

\begin{figure}[h]
\centering
\includegraphics[angle=-90,scale=0.3]{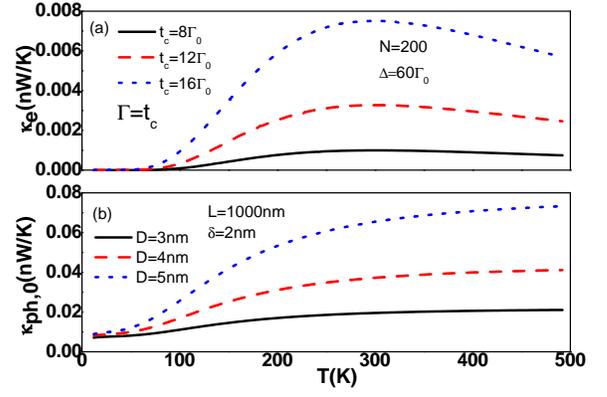}
\caption{(a) Electron thermal conductance $\kappa_e$ of SLNW with
N=200 and (b) phonon thermal conductance of nanowires with
diameters of 3, 4, and 5nm as functions of temperature.}
\end{figure}

The calculated $ZT$ as a function of temperature for various
values of $f_s$ is shown in Fig. 5. The optimum condition
$\Delta=60\Gamma_0$ has been used. It is seen in Fig. 5(a) that
$ZT$ can reach 3 at $T=300K$ with $t_c\ge 12\Gamma_0$ even with
$f_s=1$.  If we assume that phonons suffer more scattering from
QDs than electrons (i.e. $f_s<1$) as implied in previous
studies,[12,13,25] the resulting $ZT$ of SLNWs can be further
enhanced as illustrated in Fig.~5(b) for $t_c=\Gamma=16\Gamma_0$.\
The value of $ZT$ can be larger than 3 over a wide temperature
range. This feature can be very useful for the application of TEG
at room temperature.

\begin{figure}[h]
\centering
\includegraphics[angle=-90,scale=0.3]{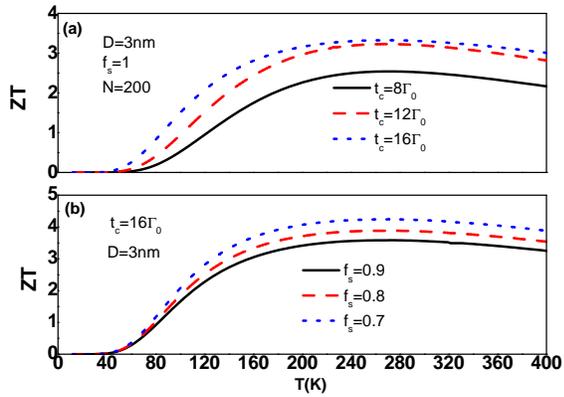}
\caption{$ZT$ as a function of temperature. (a) For different
values of $t_c (=\Gamma)$ and $f_s=1$. (b) For different values of
$f_s$ with both $t_c=\Gamma$ fixed at $16\Gamma_0$.
$\Delta=E_0-E_F=60\Gamma_0$ and $D=3~nm$.}
\end{figure}

\begin{figure}[h]
\centering
\includegraphics[angle=-90,scale=0.3]{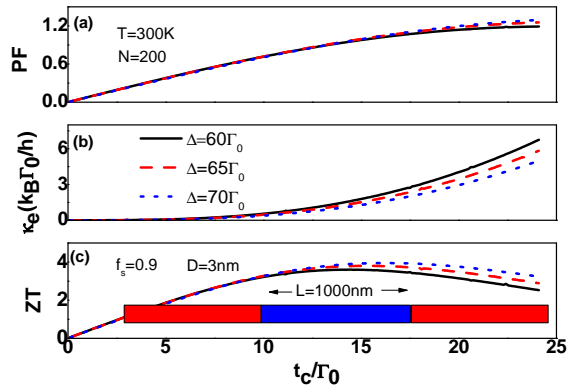}
\caption{(a) Power factor, (b) electron thermal conductance and
(c) figure of merit ($ZT$) as functions of electron hopping
strength ($t_c$) for three values of $\Delta$ at $T=300~K$ and
$N=200$. Other physical parameters are the same as those of
Fig.~5.}
\end{figure}

Finally, we examine the dependence of $ZT$ on the hopping
strength, $t_c$ (which controls the SLNW band width) for a few
values of $\Delta$ around the optimum condition.  The behaviors of
$PF$, $\kappa_e$ and $ZT$ for SLNWs  as functions of $t_c$
(=$\Gamma$) are shown in Fig. 6. It is found that $ZT$ reaches a
maximum around $t_c=\Gamma=16\Gamma_0$, while both $G_e$ and
$\kappa_e$ continue to increase with increasing $t_c(=\Gamma)$.
The rate of increase for $\kappa_e$ surpasses that for $PF$ at
$t_c=15\Gamma_0$ , leading to an optimized condition for $ZT$ at
that value. In Fig. 6(c), we found that for $t_c > 10\Gamma_0$,
the optimum condition for $\Delta$ increases slightly with the
best value occurring at $\Delta=70\Gamma_0$. The main reason for
the shift of optimum value of $\Delta$ at large $t_c$ is due to
the decrease of $\kappa_e$ for increasing $\Delta$  as seen in
Fig.~6(b).

\section{Conclusion}
We have theoretically studied the thermoelectric properties of
SLNWs connected to electrodes at room temperature.
%A linear chain model for strongly coupled QDs is used to simulate
%the electron energy levels in SLNWs parameterized by QD energy
%level $E_0$ and hopping strength, $t_c$. $E_0$ and $t_c$ as well
%as tunnel rate ($\Gamma$) are varied  in order to find the optimum
%design for SLNWs to achieve the best $ZT$ at room temperature.
%Several key characteristic results can be deduced from this model
%calculation, which can help the future design of thermoelectric
%materials based on SLNWs.
At higher temperatures (including room temperature) with $k_BT$
much greater than $t_c$ and $\Gamma$, we find the Seebeck
coefficient can be approximately described by a linear relation
$S=-\Delta/(eT)$ and the maximum of the power factor occurs at
$\Delta=2.4k_BT$. In addition, both the electrical conductance and
power factor are maximized  under the condition  $t_c=\Gamma$ for
$\Delta$ fixed around $2.4k_BT$. $ZT$ behavior is dominated by
$PF$ when $\kappa_{ph} \gg \kappa_e$. However, for large $t_c$ and
$\Gamma$, $\kappa_e$ becomes significant and the maximum $ZT$ is
determined not only by $PF=S^2G_e$, but also by $\kappa_e$. All
the above results are almost independent of the number of QDs in
the SLNW. For thin Si/Ge SLNWs with diameter around $3~nm$,  we
find that it may be possible to achieve a $ZT$ larger 3 when the
reduction of phonon thermal conductivity due to scattering from
QDs is  more severe than the reduction of electron conductivity.
The linear chain model used here did not take into account the
valley degeneracy and excited states of QDs. Adding these will
provide more channels for electron conduction, which should
improve $ZT$ further.[17]  A realistic modeling taking into
account the multi-valley band structures of Si and Ge will be left
for future investigations.

%\begin{flushleft}

%\end{flushleft}

%\begin{flushleft}
{\bf Acknowledgments}\\
%\end{flushleft}
This work was supported under Contract Nos. MOST 106-2112-M-008
-014 and MOST 104-2112-M-001-009 YM2.
\mbox{}\\
E-mail address: mtkuo@ee.ncu.edu.tw\\
E-mail address: yiachang@gate.sinica.edu.tw\\

\setcounter{section}{0}

\renewcommand{\theequation}{\mbox{A.\arabic{equation}}} %\section{Appendix}
\setcounter{equation}{0} % reset counter

%\section{}
%\subsection{Derivation of the tunneling current formula using Dyson's equations\label{App:TC_l} }
\mbox{}\\
%{\bf Appendix A. }

%\section{Appendix}

\end{document}